# Dynamics of moiré trion and its valley polarization in microfabricated WSe$_2$/MoSe$_2$ heterobilayer

*Heejun Kim[1,*], Duanfei Dong[1,*], Yuki Okamura[1,*], Keisuke Shinokita[1],*

*Kenji Watanabe[2], Takashi Taniguchi[3], and Kazunari Matsuda[1]*

[1]Institute of Advanced Energy, Kyoto University, Uji, Kyoto 611-0011, Japan

[2]Research Center for Functional Materials, National Institute for Materials Science, 1-1 Namiki, Tsukuba, Ibaraki 305-0044, Japan

[3]International Center for Materials Nanoarchitectonics, National Institute for Materials Science, 1-1 Namiki, Tsukuba, Ibaraki 305-0044, Japan

**Abstract**

The moiré potential, induced by stacking two monolayer semiconductors with slightly different lattice mismatches, acts as periodic quantum confinement for optically generated excitons, resulting in spatially ordered zero-dimensional quantum systems. However, there are limitations to exploring intrinsic optical properties of moiré excitons due to ensemble-averaged and broadened emissions from many peaks caused by the inhomogeneity of the moiré potential. In this study, we proposed a microfabrication technique based on focused $Ga^+$ ion beams, which enables us to control the number of peaks originating from the moiré potential and thus explore unknown moiré optical characteristics of $WSe_2/MoSe_2$ heterobilayers. By taking advantage of this approach, we reveal emissions from a single moiré exciton and charged moiré exciton (trion) under electrostatic doping conditions. We show the momentum dark moiré trion state above the bright trion state with a splitting energy of approximately 4 meV and clarify that the dynamics are determined by the initial trion population in the bright state. Furthermore, the degree of negative circularly polarized emissions and their valley dynamics of moiré trions are dominated by a very long valley relaxation process lasting ~700 ns. Our findings on microfabricated heterobilayers could be viewed as an extension of our groundbreaking efforts in the field of quantum optics application using moiré superlattices.

**Introduction**

The development of optical physics of tightly bound electron–hole pairs (exciton) in atomically thin two-dimensional (2D) materials, such as semiconducting monolayer transition metal dichalcogenides, has significant considerable interest in recent years.[1–4] Moiré superlattices with varying atomic registries can be constructed by stacking two monolayers with a small lattice mismatch or twist angle, which drastically modulates the electronic band structure of monolayer semiconductors.[5–10] The resultant moiré pattern leads to the formation of periodic ordered potential traps, which confine and spatially organize excitons as quantum two-level system ensembles.[11–14] The trapped exciton (moiré exciton) exhibits a narrow emission peak (<1 meV) that selectively couples to circularly polarized light on $C_3$ rotational symmetry,[15–20] providing a novel platform for developing programmable quantum optics applications.[21–24] However, investigating the intrinsic optical properties arising from the moiré excitonic state, including moiré excitons due to broadened and ensemble-averaged emissions composed of many peaks, is challenging. These broad emissions can be mainly attributed to the inhomogeneity of the moiré potential.[15,25–27]

In the addition to moiré excitons, research on charged moiré excitons (moiré trions) as a bound state of moiré excitons and charge carriers within the moiré potential in twisted semiconducting heterobilayers, has rapidly progressed. Several studies have been conducted to clarify the emergence of moiré trion peaks at the lower energy side of neutral moiré excitons and circularly polarized light emissions.[28–31] Although some studies have characterized the intrinsic properties of moiré trions, their valley dynamics between discrete energy states within the moiré potential remain elusive. Thus, more studies are required to reveal the excitonic structures of moiré trions within the moiré potential, which is hindered by inhomogeneously broadened optical spectra in semiconducting heterobilayers. Such studies could be strongly considered as an extension of groundbreaking efforts in the field of fundamental moiré physics and potential quantum optics applications toward a more challenging and unexplored domain of moiré excitonic systems.

In this study, we propose a novel microfabrication technique based on $Ga^+$ ion beams for twisted 2D semiconducting van der Waals (vdW) heterobilayers. The proposed technique allows us to control the optical excitation and detection area while using a limited number of moiré potentials without being constrained by the diffraction limit of light. This novel approach offers significant clear optical spectra composed of only few peaks by overcoming inhomogeneously broadened and ensemble-averaged spectra resulting from moiré excitonic states. We have revealed the fine structures of moiré trion states and their valley dynamics by taking advantage of the microfabricated $WSe_2$/$MoSe_2$ heterobilayer.

## 2. Results and Discussion

**Figure 1a** shows a schematic of the microfabricated $WSe_2$/$MoSe_2$ heterobilayer with top and bottom *h*-BN capping layers (sample A) used in this study, which is prepared by the dry-transfer technique using the polymer stamp method **(Supporting Information)**. During optical measurements, it is difficult to obtain clear and significant spectra from moiré excitonic states due to the inhomogeneity of the moiré potential, resulting in ensemble-averaged and broadened emissions composed of multiple peaks,[15,32,33] as schematically shown in Figure 1a. To address this problem, in the microfabrication process, we first used the focused Ga$^+$ ion beam (FIB) technique to reduce the optical excitation and detection area of the WSe$_2$/MoSe$_2$ heterobilayer with a micro-pillar and bridged structure. **Figure 1b** shows a typical optical image of the microfabricated WSe$_2$/MoSe$_2$ heterobilayer, where the areas surrounded by the red and blue solid lines correspond to the monolayer WSe$_2$ and MoSe$_2$ regions, respectively. The twist angle between two monolayers can be estimated as ~ 1.5° from the alignment of the monolayer edges of MoSe$_2$ and WSe$_2$ by optical microscope image. The twist angle was further confirmed by the second-harmonic generation (SHG) measurement, which is consistent with the estimated angle from optical microscope image. (**Figure S1, See Supporting Information**). **The inset of Figure 1b** shows a typical optical image of the fabricated microstructure arrays in the heterobilayer, with different pillar diameter sizes *D*, ranging from 100 nm to 1.5 μm, using the FIB technique.

**Figure 1c** shows a schematic of the optical measurements of the WSe$_2$/MoSe$_2$ heterobilayer without and with microfabrication using focused Ga$^+$ ion beams. The focused laser light with a spot size of ~1.5 μm is determined by the diffraction limit of light, which excites large number of moiré potentials because of the significantly smaller spatial period of moiré potentials in the WSe$_2$/MoSe$_2$ heterobilayer. This might cause very broad photoluminescence (PL) signals caused by radiative recombination of moiré excitons from large number of moiré potentials of WSe$_2$/MoSe$_2$ heterobilayers without microstructures (Figure 1c, left). In contrast, significantly smaller optical excitation and detection determined by the physical pillar size of *D*, not limited by the diffraction limit of light, are realized in the microfabricated WSe$_2$/MoSe$_2$ heterobilayer. The microfabricated WSe$_2$/MoSe$_2$ heterobilayer with a smaller pillar size ($D <$ ~400 nm) provides a reduced number of PL peaks while avoiding broadened and ensemble-averaged spectra caused by the inhomogeneity of the moiré potential beyond the diffraction limit of light (Figure 1c, right).

**Figure 1d** shows typical PL spectra of the microfabricated $WSe_2/MoSe_2$ heterobilayer with various pillar sizes (1 µm, 400 and 200 nm). The PL spectra of the $WSe_2/MoSe_2$ heterobilayer with a larger pillar size ($D = 1$ µm) exhibit multiple inhomogeneously broadened peaks. A significant decrease in the number of peaks in the inhomogeneously broadened PL spectra is observed with decreasing pillar size. Then, sharp emission features with a few spectral peaks are realized in the $WSe_2/MoSe_2$ heterobilayer with a smaller pillar size ($D = 200$ nm), with the expectation that PL signals will be detected from a reduced finite number of moiré potentials by the microfabrication technique. With the advances in the microfabrication technique to reduce the number of peaks caused by the inhomogeneity of the moiré potential, we can further investigate additional unexplored optical characteristics of the sample **(Figure S2, Supporting Information**).

Taking advantage of the microfabrication process to reduce the number of peaks caused by the inhomogeneity of the moiré potential, we explore charged excitons (trions) from microfabricated $WSe_2/MoSe_2$ heterobilayer with a graphene electrode (sample B) shown in **Figure S3, Supporting Information**. **Figure 2a** shows the PL spectrum of the microfabricated $WSe_2/MoSe_2$ heterobilayer ($D \approx 250$ nm) by varying the applied back-gate voltage ($V_g$) from −30 to 30 V at 8.5 K. The small number of multiple narrow PL peaks arising from moiré excitons is observed in the range of 1.33–1.36 eV at $V_g$ of −20 V due to the microfabricated WSe$_2$/MoSe$_2$ heterobilayer. With an increase in $V_g$ across 0 V, narrow moiré exciton PL peaks vanish, and a new set of peaks emerges at the lower energy side. By further increasing $V_g$ up to 30 V, PL peaks are further redshifted and broadened with decreasing intensity. These subsequent energy position changes indicate the charging effects of the back-gate voltage on moiré excitons, resulting from the formation of bound states of moiré excitons and charge carriers as charged moiré excitons (moiré trions) in the same moiré potential traps.[29,34] **(Figure S4, Supporting Information)**

These changes are observed in the contour map of back-gate voltage-dependent PL spectra with doped carrier density ($n_e$), as shown in **Figure 2b**, where the doped carrier density is evaluated using the capacitance model with consideration of a slight shift of charge neutral point (See the **Supporting Information**). Significant spectral changes are observed at a doped carrier density ($n_e$) of $2 \times 10^{12}$ cm$^{-2}$ in the counter map. The doped carrier density of $2 \times 10^{12}$ cm$^{-2}$ well corresponds to the number density of the moiré potential ($D \approx 2 \times 10^{12}$ cm$^{-2}$) calculated from a period of moiré superlattices (6.2 nm) in the WSe$_2$/MoSe$_2$ heterobilayer at a twist angle of ~3° from the optical image in Figure S3b (See the **Supporting Information**), which suggests that the spectral change occurs under the condition of a doped electron per a moiré potential. **The inset of Figure S4** shows the PL spectra of neutral moiré excitons ($M_X^0$) and moiré trions ($M_T^-$), where the PL spectrum of $M_T^-$ is shifted to a lower energy side of 7.3 meV. The PL spectra of $M_X^0$, and $M_T^-$ has a similar shape, which suggests that each PL peak at the same energy level originated from the same moiré potential. The binding energy of moiré trions is estimated as approximately 7 meV from the energy difference between peaks of $M_X^0$ and $M_T^-$ in the same moiré potential, which is consistent with the previously reported results.[28,29,31]

**Figure 2c** shows a plot of integrated PL intensities as a function of the doped carrier density from the counter map. The PL intensity monitored at 1.337 and 1.347 eV shows a maximum value around the neutral condition ($n_e \ll 2\times10^{12}$ cm$^{-2}$), whose peaks are attributed to the recombination of neutral moiré excitons ($M_{X_1}^0$ and $M_{X_1}^0$). Emergent discrete several PL peaks in the lower energy region from neutral moiré exciton peaks exhibit the maximum intensity at the slight electron-doping regime ($n_e < 2 \times 10^{12}$ cm$^{-2}$), as shown in Figure 2b, which are assigned as charged moiré excitons or trions ($M_T^-$), where neural moiré excitons interact with a doped electron in the moiré potential to form a moiré trion. Moreover, further spectral changes are observed at the high electron-doping regime ($n_e > 2 \times 10^{12}$ cm$^{-2}$) in the counter map in Figure 2b, where the two electrons per moiré potential are electrostatically doped. The PL intensity monitored at 1.323 and 1.328 eV shows a maximum value under high electron-doping conditions ($n_e \gg 2 \times 10^{12}$ cm$^{-2}$), whose peak may have resulted from the recombination of the bound state of a moiré trion and an electron ($M_{T_1}^{--}$ and $M_{T_2}^{--}$). We further measured the excitation power dependence of PL intensities of neutral moiré excitons and moiré trions under $V_g$ of 10 and -20 V, as shown in **Figure S5, Supporting Information**. The strong nonlinear saturation behavior and its corresponding saturation density of $2 \times 10^{12}$ cm$^{-2}$ also support the signature of the moiré trions confined in the moiré potential.[35,36]

**Figure 3a** shows a contour plot of temperature-dependent PL spectra of moiré trions from 8.5 to 30 K at a $V_g$ of 10 V. The **inset of Figure 3a** shows the monitored PL intensities of the four narrow peaks (1.324, 1.329, 1.334, and 1.341 eV) from moiré trions as a function of temperature from 8.5 to 30 K. The PL intensities of all peaks exhibit rapid quenching upon increasing temperature, which shows Arrhenius-type behavior. The activation energy evaluated from the four peaks is estimated to be approximately 4 meV, which is smaller than that of a moiré exciton of ~8 meV and the moiré potential depth for moiré excitonic states of approximately 150 meV.[8,15,37] Moreover, the evaluated larger moiré trion binding energy of approximately 7 meV suggests that the moiré trion is confined to the same potential as the moiré exciton. These findings strongly indicate that thermal quenching of PL intensity by moiré trions does not originate from the thermal activation of excitons outside the moiré potential, but rather from the additional dark state that does not contribute to the emission inside the moiré potential, as discussed later.[15,36,38]

The temperature dependence of the moiré trion dynamic further evokes the presence of additional energy levels within the moiré potential. **Figure 3b** shows the temperature dependence of the PL decay profiles of interlayer moiré trions under $V_g$ of 10 V. We confirm that the temperature dependence of integrated PL intensity evaluated from the time-integrated PL decay profiles from 5 to 30 K is well consistent with that of steady-state PL intensity, as illustrated in **Figure S6, Supporting Information**. The PL decay curves of moiré trions show a longer lifetime of several tenth ns, suggesting a smaller oscillator strength of the interlayer moiré trion, similar to that of the interlayer moiré exciton.[3,4,21,28,39,40] The PL decay curve remains almost constant until 20 K and only decreases slightly at 30 K, which is inconsistent with the decrease in steady-state PL intensity with increasing temperature shown in Figure 3a, as discussed later.

The temperature dependence of the PL decay curves was fitted by bi-exponential functions of $I(t) = A_1 \exp\left(-\frac{t}{\tau_f}\right) + A_2 \exp(-\frac{t}{\tau_s})$, where $A_{1(2)}$ and $\tau_{f(s)}$ denote the coefficients and decay times from the fast (slow) components, respectively.[6,7,21] The solid black curve represents the fitted results from a bi-exponential function, which reproduces the PL decay curves well over the entire temperature range, as shown in the **inset of Figure 3b.** The PL decay times are almost constant until 20 K, which is inconsistent with a decrease in steady-state PL intensity with increasing temperature. These inconsistencies between the decrease in PL intensity with increasing temperature and almost constant decay lifetime at a high temperature (>20 K) suggest that the temperature-dependent dynamics of the initial trion population in the bright and dark states play an important role.

For further analysis of the decay dynamics of the initial trion population from bi-exponential functions, we model a three-level energy system with dark, bright, and ground states using rate equations shown in **Figure S7a, Supporting Information**. The phenomenological rate equation, including decay parameters of the bright moiré trion radiative decay rate ($\tau_{Br}^{-1}$), non-radiative dark trion decay rate ($\tau_{Dnr}^{-1}$), and transition rate between the bright and dark states ($\tau_{BD}^{-1}$ and $\tau_{DB}^{-1}$), can be described by the following equations:

$$\frac{dN_B(t)}{dt} = -N_B(t)(\tau_{Br}^{-1} + \tau_{BD}^{-1}) + N_D(t)\tau_{DB}^{-1}, \tag{1}$$

$$\frac{dN_D(t)}{dt} = -N_D(t)(\tau_{Dnr}^{-1} + \tau_{DB}^{-1}) + N_B(t)\tau_{BD}^{-1}, \tag{2}$$

where $N_B(t)$ and $N_D(t)$ denote temperature-dependent trion populations in the bright and dark states, respectively. We assumed that the initial trion populations in the bright and dark states are significantly affected by the trion generation of $G$, where $G$ $(=N_B(0)/(N_B(0)+N_D(0)))$ at different temperatures, as discussed later. The phonon-assisted transition ($\tau_{BD}^{-1}(T)$ and $\tau_{DB}^{-1}(T)$), can be described by the phonon absorption process ($\tau_{BD}^{-1}(T) = \tau_{0(BD)}^{-1}(0) <n>$) and phonon emission process ($\tau_{DB}^{-1}(T) = \tau_{0(DB)}^{-1}(0)(<n+1>)$), where $<n> = 1/[\exp(\Delta E_{BD}/k_B T)-1]$ denotes the phonon occupation number at temperature $T$.[41,42]

As a negligible number of phonons ($<n> \cong 0$) with a larger splitting energy of the bright and dark trion states ($k_B T \ll \Delta E_{BD}$) is considered at a low-temperature (5 K), the fast and slow decay rates can be expressed as $\tau_f^{-1} \cong \tau_{Br}^{-1} + \tau_{BD}^{-1}$ and $\tau_s^{-1} \cong \tau_{Dnr}^{-1} + \tau_{DB}^{-1}$, respectively, indicating the population decay in the moiré bright state $N_B(t)$. **Figure S7b** shows the temperature-dependent decay times derived from the experimental results of the PL decay profiles shown in the inset of Figure 3b. With an increase in the temperature up to 20 K, the phonon-assisted transition rate ($\tau_{BD}^{-1}$) accelerates, but it is still significantly slower than the radiative lifetime of the bright moiré trion state ($\tau_{Br}^{-1}$). The relationship between $\tau_{Br}^{-1}$ and $\tau_{BD}^{-1}$ can be used to illustrate the PL decays derived from the bi-exponential function in Figure 3b, which indicates the constant fast and slow decay behavior until 20 K. When the temperature exceeds 30 K, $\tau_{BD}^{-1}$ becomes faster than $\tau_{Br}^{-1}$ under the condition ($k_B T > \Delta E_{BD}$). Then, significant decreases in lifetime for both $\tau_f$ and $\tau_s$ are observed.

To confirm the validity of the rate equation analysis based on the initial trion population in the bright and dark energy states, we analyzed amplitudes, $A_1$ and $A_2$ (**Figure S7c, Supporting Information**) and the steady-state PL intensity as a function of temperature using Eqs. (3) and (4), as follows,[43,44]

$$A_{1(2)} = \pm N_B(0) \frac{\tau_{Br}^{-1} + \tau_{BD}^{-1} + \tau_{Dnr}^{-1} - \tau_{s(f)}^{-1}}{\tau_f^{-1} - \tau_s^{-1}} \mp N_D(0) \frac{\tau_{DB}^{-1}}{\tau_f^{-1} - \tau_s^{-1}}, \quad (3)$$

$$I(T) \propto \tau_{rad}^{-1} \frac{G}{(\tau_{Br}^{-1} + \tau_{BD}^{-1}) - \left\{ \frac{\tau_{DB}^{-1} \tau_{BD}^{-1}}{(\tau_{Dnr}^{-1} + \tau_{DB}^{-1})} \right\}}. \quad (4)$$

Notably, the values of the amplitudes ($A_1$ and $A_2$) with the initial bright $N_B(0)$ and dark $N_D(0)$ populations using Eq. (3) are well reproduced with the amplitude from the bi-exponential functions shown **in Figure 3c**. In addition, the temperature dependence of the derived PL intensity using Eq. (4) further demonstrates that the PL intensity is determined by the initial trion generation of $G$, where $G$ ($=N_B(0)/(N_B(0)+N_D(0))$) is defined. These results strongly suggest that the initial population of bright moiré trions decreases drastically and that many moiré trions initially populate in the higher state at high temperatures, which determines the moiré trion dynamics in the moiré potential. Then, the non-radiative recombination of trions in the dark state drastically decreases PL intensities at high temperatures,

We discuss the physical origin of the dark moiré trion state, as shown in **Figure 3d**. Given the energy difference of approximately 4 meV between the bright trion state and the *ab-initio* calculation,[45–47] the dark moiré trion may have originated from the momentum-related dark states, where the electrons in the conduction band of K (K⁻) valley might scatter to the different valleys of Q or I- due to phonon transitions with increasing temperature. Thus, the moiré dark trion can be attributed to the K-K exciton interacting with excess electrons from the Q(I-) valley trapped in the moiré potential, which also contributes to anomalous valley polarization dynamics, as described later.

**Figure 4a** shows the polarization-resolved PL spectra of moiré trions at various temperatures under $\sigma^+$ excitation. The excitation photon energy of 1.72 eV corresponds to the resonant excitation of excitons at the K valley in the WSe$_2$ layer of the heterobilayer. The red and blue curves show the PL spectra of moiré trions monitored at $\sigma^+$ and $\sigma^-$ circularly polarized components, respectively. The PL peak at 1.329 eV shows a stronger $\sigma^-$ component than $\sigma^+$ component, indicating negative circular polarization originating from the optical selection rule, which is consistent with a previous report.[16,18] As the temperature increases, the PL peak at 1.329 eV shows a similar intensity $\sigma^+$ and $\sigma^-$ emission and disappearance of negative circular polarization at 30 K.

The time evolution of circular polarization also provides important information about the valley dynamics of moiré trions. **Figure 4b** shows the polarization-resolved PL decays of the $\sigma^+$ and $\sigma^-$ components at 1.329 eV under $\sigma^+$ pulse excitation at various temperatures. The intensity of the $\sigma^-$ component in time evaluation is stronger than that of the $\sigma^+$ component from 5 to 15 K and becomes similar to that of the $\sigma^+$ component above 30 K. The trend of temperature dependent polarization-resolved PL decay is further confirmed at various temperatures in the range of 5–30 K, as shown in **Figure S8a, Supporting Information**. These results are well consistent with the degrees of circular polarization obtained in the polarization-resolved PL spectra, as depicted in Figure 4a. **Figure 4c** shows the degree of circular polarization $\rho$ as a function of temperature from polarization-resolved PL spectra in Figure 4a, defined as $\rho = \frac{I_{\sigma^+} - I_{\sigma^-}}{I_{\sigma^+} + I_{\sigma^-}}$, where $I_{\sigma^+}$ and $I_{\sigma^-}$ denote the PL intensities of the $\sigma^+$ and $\sigma^-$ components, respectively. The degree of polarization at low temperatures shows large negative values and decreases with increasing temperature.

**Figure 4d** demonstrates the time evolution of valley polarization $\rho(t) = \frac{I_{\sigma^+}(t) - I_{\sigma^-}(t)}{I_{\sigma^+}(t) + I_{\sigma^-}(t)}$, from the polarization-resolved PL decay curves of $I_{\sigma^+}(t)$ and $I_{\sigma^-}(t)$ in Figure 4b, as a function of temperature. The temperature dependence of the time-integrated valley polarization $\rho(t)$ is well matched with that of the steady-state valley polarization $\rho$, as depicted in Figure 4c, which shows the consistency of the steady-state and time-resolved optical experiments **(Figure S8b, Supporting Information)**. The evaluated time evolution of the negative valley polarization shows a very long lifetime of approximately 700 ns, corresponding to the valley relaxation lifetime of a bright moiré trion even as the temperature increases, and is significantly longer than the radiative lifetime of several tenths of ns **(Figure S9, Supporting Information)**. The valley polarization is phenomenologically described as $\rho = \frac{\rho(0)}{1 + 2\tau_r/\tau_v}$,[48,49] where $\rho(0)$ denotes the initial valley polarization. The valley polarization of a bright moiré trion is mainly determined by the initial valley polarization $\rho(0)$ under the condition of a very long and temperature-invariant valley relaxation lifetime. Thus, the decrease in the valley polarization with increasing temperature in Figure 4a can be explained by temperature-dependent moiré trion dynamics, as discussed above.

**Figure 4e** shows a schematic of the bright and dark moiré trion configuration and temperature-dependent moiré trion dynamics, where the size of the black circle reflects the population of electrons and holes comprising the moiré trion states. At low temperatures below 20 K, the radiative recombination of dominant bright moiré trions composed of electrons at K (K') valley and holes at the K valley generated under $\sigma^+$ excitation shows $\sigma^-$ emission in accordance with the valley selection rule. In contrast, the initial population transfer from bright to dark moiré trions composed of electrons at the Q(I-) and K' valleys and holes at K valley with increasing temperature results in the disappearance of $\sigma^-$ emission dominated by the bright moiré trion. The bright and dark moiré trion states and their moiré trion dynamics studied in this study reveal the intriguing temperature-dependent initial trion populations in the bright and dark states and circularly polarized emission of moiré trions with a very long temperature-invariant valley relaxation process.

**Conclusion**

We developed an advanced optical technique to obtain significant PL spectra with a few spectral peaks from the moiré excitonic states in the microfabricated $WSe_2/MoSe_2$ heterobilayer. This approach offers an opportunity to unveil intrinsic moiré optical characteristics beyond the diffraction limit of light, which enables the study of the fine structures of the moiré trion states and their intriguing valley dynamics in the moiré potential. The rapid decreases in the experimentally obtained PL intensity with increasing temperature reflect the presence of a momentum-related dark moiré trion state lying above the bright moiré trion state with an energy splitting of 4 meV within the moiré potential traps. The temperature dependence of time-resolved PL spectroscopy and detail analysis using the rate equations revealed that the dynamics of moiré trions are primarily determined by the initial population in the bright state at low temperatures and are highly populated in the dark state at high temperatures. Moreover, the intriguing temperature-dependent initial trion population results in the negative circularly polarized emission of moiré trions with a very long temperature-invariant valley relaxation process lasting 700 ns. Our experimental results reflect new aspects in exploring the quantum states of moiré trions and may be seen as an extension of our groundbreaking efforts regarding the understanding of quantum phenomena in moiré physics for quantum optics applications.

**Reference**

1. Luo, R. *et al.* Van der Waals interfacial reconstruction in monolayer transition-metal dichalcogenides and gold heterojunctions. *Nat. Commun.* **11**, 1011 (2020).

2. Munkhbat, B. *et al.* Transition metal dichalcogenide metamaterials with atomic precision. *Nat. Commun.* **11**, 4604 (2020).

3. Miller, B. *et al.* Long-Lived Direct and Indirect Interlayer Excitons in van der Waals Heterostructures. *Nano Lett.* **17**, 5229–5237 (2017).

4. Rivera, P. *et al.* Observation of long-lived interlayer excitons in monolayer $MoSe_2$-$WSe_2$ heterostructures. *Nat. Commun.* **6**, 6242 (2015).

5. Yuan, L. *et al.* Twist-angle-dependent interlayer exciton diffusion in $WS_2$–$WSe_2$ heterobilayer. *Nat. Mater.* **19**, 617–623 (2020).

6. Choi, J. *et al.* Twist Angle-Dependent Interlayer Exciton Lifetimes in van der Waals Heterostructures. *Phys. Rev. Lett.* **126**, 047401 (2021).

7. Nayak, P. K. *et al.* Probing Evolution of Twist-Angle-Dependent Interlayer Excitons in $MoSe_2/WSe_2$ van der Waals Heterostructures. *ACS Nano.* **11**, 4041–4050 (2017).

8. Yu, H. *et al.* Moiré excitons: From programmable quantum emitter arrays to spin-orbit-coupled artificial lattices. *Sci. Adv.* **3**, e1701696 (2017).

9. Alexeev, E. M. *et al.* Resonantly hybridized excitons in moiré superlattices in van der Waals heterostructures. *Nature.* **567**, 81–86 (2019).

10. Andersen, T. I. *et al.* Excitons in a reconstructed moiré potential in twisted $WSe_2/WSe_2$ homobilayers. *Nat. Mater.* **20**, 480–487 (2021).

11. Interlayer excitons and how to trap them. *Nat. Mater.* **19**, 579 (2020).

12. Seyler, K. L. *et al.* Signatures of moiré-trapped valley excitons in $MoSe_2/WSe_2$ heterobilayer. *Nature.* **567**, 66–70 (2019).

13. Tartakovskii, A. Moiré or not. *Nat. Mater.* **19**, 581–582 (2020).

14. Jin, C. *et al.* Observation of moiré excitons in $WSe_2/WS_2$ heterostructure superlattices. *Nature.* **567**, 76–80 (2019).

15. Tran, K. *et al.* Evidence for moiré excitons in van der Waals heterostructures. *Nature.* **567**, 71–75 (2019).

16. Shinokita, K. *et al.* Valley Relaxation of the Moiré Excitons in a $WSe_2/MoSe_2$ Heterobilayer. *ACS Nano.* **16**, 16862–16868 (2022).

17. Zhang, L. *et al.* Highly valley-polarized singlet and triplet interlayer excitons in van der Waals heterostructure. *Phys. Rev. B.* **100**, 041402 (2019).

18. Hsu, W. T. *et al.* Negative circular polarization emissions from $WSe_2/MoSe_2$ commensurate heterobilayer. *Nat. Commun*. **9**, 1356 (2018).

19. Ciarrocchi, A. *et al.* Polarization switching and electrical control of interlayer excitons in two-dimensional van der Waals heterostructures. *Nat. Photon*. **13**, 131–136 (2019).

20. Rivera, P. *et al.* Interlayer valley excitons in heterobilayer of transition metal dichalcogenides. *Nat. Nanotechnol*. **13**, 1004–1015 (2018).

21. Li, Z. *et al.* Interlayer Exciton Transport in $MoSe_2/WSe_2$ Heterostructures. *ACS Nano*. **15**, 1539–1547 (2021).

22. Jiang, Y, *et al*. Interlayer exciton formation, relaxation, and transport in TMD van der Waals heterostructures. *Light Sci. Appl.* **10**, 72 (2021).

23. Wu, B. *et al.* Observation of moiré excitons in the twisted $WS_2/WS_2$ homostructure. *Nanoscale*. **14**, 12447–12454 (2022).

24. Camacho-Guardian, A. *et al*. Moiré-Induced Optical Nonlinearities: Single- and Multiphoton Resonances. *Phys. Rev. Lett*. **128**, 207401 (2022).

25. Guo, H., *et al*. Shedding light on moiré excitons: A first-principles perspective. *Sci. Adv*. **6**, 42 (2020).

26. Susarla, S. *et al.* Hyperspectral imaging of exciton confinement within a moiré unit cell with a subnanometer electron probe. *Science*. **378** 1235-1239 (2022)

27. Karni, O. *et al.* Structure of the moiré exciton captured by imaging its electron and hole. *Nature*. **603**, 247–252 (2022).

28. Wang, X. *et al.* Moiré trions in $MoSe_2/WSe_2$ heterobilayer. *Nat*. *Nanotechnol*. **16**, 1208–1213 (2021).

29. Liu, E. *et al.* Signatures of moiré trions in WSe$_2$/MoSe$_2$ heterobilayer. *Nature*. **594**, 46–50 (2021).

30. Calman, E. v. *et al.* Indirect Excitons and Trions in MoSe$_2$/WSe$_2$ van der Waals Heterostructures. *Nano Lett*. **20**, 1869–1875 (2020).

31. Dandu, M. *et al.* Electrically Tunable Localized versus Delocalized Intralayer Moiré Excitons and Trions in a Twisted MoS$_2$ Bilayer. *ACS Nano*. **16**, 8983–8992 (2022).

32. Zheng, H. *et al.* Evidence for interlayer coupling and moiré excitons in twisted WS$_2$/WS$_2$ homostructure superlattices. *Nano*. *Res*. (2022) doi:10.1007/s12274-022-4964-4.

33. Baek, H. *et al.* Highly energy-tunable quantum light from moiré-trapped excitons. *Sci. Adv*. **6**, eaba8526 (2020).

34. Baek, H. *et al.* Optical read-out of Coulomb staircases in a moiré superlattice via trapped interlayer trions. *Nat*. *Nanotechnol*. **16**, 1237–1243 (2021).

35. Shinokita, K. *et. al*. Resonant Coupling of a Moiré Exciton to a Phonon in a WSe$_2$/MoSe$_2$ Heterobilayer. *Nano Lett*. **21**, 5938–5944 (2021).

36. Zhang, Y. *et al.* Magnon-Coupled Intralayer Moiré Trion in Monolayer Semiconductor–Antiferromagnet Heterostructures. *Adv*. *Mater*. **34**, e2200301 (2022).

37. Lu, X. *et al*. Modulated interlayer exciton properties in a two-dimensional moiré crystal. *Phys*. *Rev*. *B*. **100**, 155416 (2019).

38. Yuan, Q. *et al.* Interplay Effect of Temperature and Excitation Intensity on the Photoluminescence Characteristics of InGaAs/GaAs Surface Quantum Dots. *Nanoscale*. *Res*. *Lett*. **13**, 387 (2018).

39. Choi, J. *et al.* Moiré potential impedes interlayer exciton diffusion in van der Waals heterostructures. *Sci. Adv*. **6** eaba8866 (2020).

40. Wibmer, L. *et al*. Excitons and Trions in One-Photon- and Two-Photon-Excited $MoS_2$: A Study in Dispersions. *Adv. Mater.* **30**, 1706702 (2018).

41. Shinokita, K. *et al.* Phonon-mediated intervalley relaxation of positive trions in monolayer $WSe_2$. *Phys. Rev. B*. **100**, 161304 (2019).

42. Miyauchi, Y. *et al.* Evidence for line width and carrier screening effects on excitonic valley relaxation in 2D semiconductors. *Nat. Commun*. **9**, 2598 (2018).

43. Dalgarno, P. A. *et al.* Dark exciton decay dynamics of a semiconductor quantum dot. *Phys. Status. Solidi (A). Appl*. **202**, 2591–2597 (2005).

44. Wang, G. *et al.* Voltage control of the spin dynamics of an exciton in a semiconductor quantum dot. *Phys. Rev. Lett*. **94**, 197402 (2010).

45. Förg, M. *et al.* Moiré excitons in $MoSe_2$-$WSe_2$ heterobilayer and heterotrilayers. *Nat Commun*. **12**, 1656 (2021).

46. Heo, H. *et al.* Interlayer orientation-dependent light absorption and emission in monolayer semiconductor stacks. *Nat Commun*. **6**, 7372 (2015).

47. Nayak, P. K. *et al.* Probing Evolution of Twist-Angle-Dependent Interlayer Excitons in $MoSe_2$/$WSe_2$ van der Waals Heterostructures. *ACS Nano*. **11**, 4041–4050 (2017).

48. Zhang, Y. *et al.* Magnetic Field Induced Inter-Valley Trion Dynamics in Monolayer 2D Semiconductor. *Adv. Funct. Mater*. **31**, 2006064 (2021).

49. Shinokita, K. *et al.* Continuous Control and Enhancement of Excitonic Valley Polarization in Monolayer $WSe_2$ by Electrostatic Doping. *Adv. Funct. Mater*. **29**, 1900260 (2019).

ASSOCIATED CONTENT

**Supporting Information**.

The following files are available free of charge.

Measurement of twist angle in a twisted WSe$_2$/MoSe$_2$ heterobilayer; Measurement of the charge carrier doping using capacitance model; Estimation of the exciton density in moiré potential by lattice constant; Steady-state PL intensity as a function of temperature based on rate equation analysis; Stacking angle of WSe$_2$/MoSe$_2$ heterobilayer using SHG measurement; Optical characteristics from micro-fabricated WSe$_2$/MoSe$_2$ heterobilayer; Optical characteristics from micro-fabricated WSe$_2$/MoSe$_2$ heterobilayer with a graphene electrode; PL spectra of WSe$_2$/MoSe$_2$ heterobilayer at different back-gate voltage conditions; Power dependent PL spectra of WSe$_2$/MoSe$_2$ heterobilayer at different $V_g$; PL decay curves of WSe$_2$/MoSe$_2$ heterobilayer at various temperatures; Analysis of temperature dependence of lifetimes and relative amplitudes of PL decay curves; Temperature dependence of polarization resolved PL decays and transient degree of valley polarization; Temperature dependence of valley polarization from polarization-resolved PL decay curves

AUTHOR INFORMATION

**Corresponding Author**

Kazunari Matsuda: Institute of Advanced Energy, Kyoto University, Uji, Kyoto 611-0011, Japan; Email: matsuda@iae.kyoto-u.ac.jp

**Author Contribution**

D.D., Y.O., K.W., and T.T. contributed to the fabrication of samples studied here. H.K., K.S., and K.M. designed the experiments that were performed by H.K., D.D. and Y.O. Data analysis was performed by H.K., D.D., and K.M. The draft was written by H.K.,Y.O., D.D., and K.M., with all authors contributing to reviewing and editing. The project was supervised by K.M.

**Notes**

Authors declare that they have no competing interests.

**Acknowledgements**

This work was supported by JSPS KAKENHI (Grant Numbers JP16H06331, JP19K14633, JP19K22142, JP20H05664, JP21H05232, JP21H05235, JP21H01012, JP21H05233, and JP22K18986), JST FOREST program (Grant Number JPMJFR213K), and the Collaboration Program of the Laboratory for Complex Energy Processes, Institute of Advanced Energy, Kyoto University. Growth of $h$-BN was supported by the JSPS KAKENHI (Grant Numbers JP19H05790, JP20H00354, and JP21H05233).

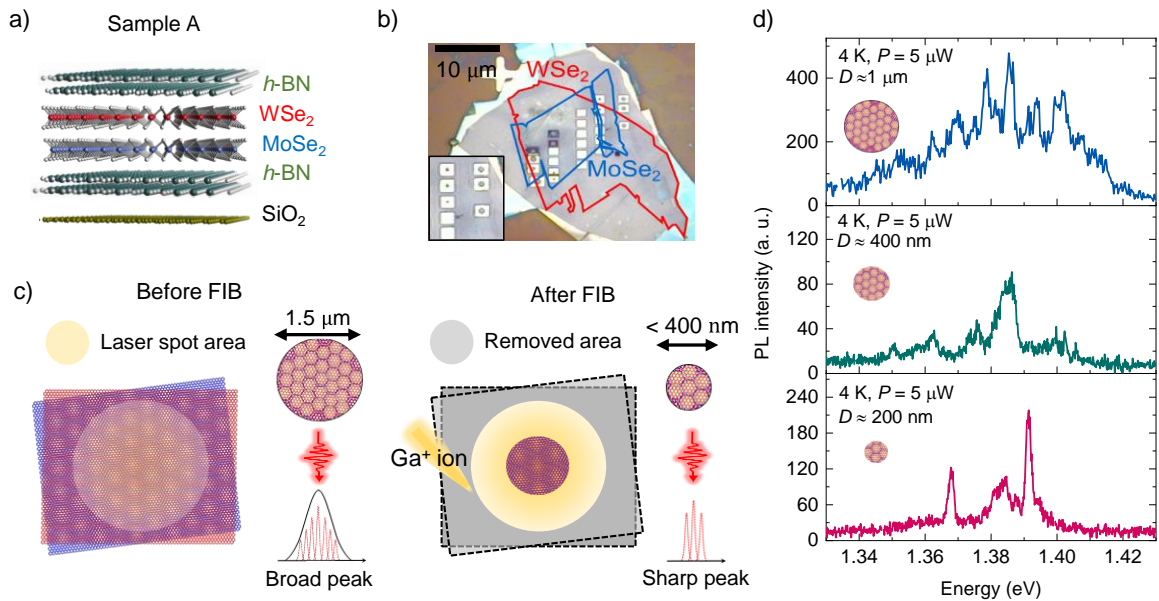

**Figure 1.** (a) Schematic of 1L-WSe$_2$/MoSe$_2$ vdW heterobilayer with top and bottom of *h*-BN. (b) Optical image of microfabricated WSe$_2$/MoSe$_2$ heterobilayer, where the areas surrounded by the red and blue solid lines correspond to monolayer WSe$_2$ and MoSe$_2$ regions, respectively. The scale bar of 10 μm is shown in the image. Inset shows a typical image of microfabrication using the FIB technique. (c) Schematic of optical measurements of WSe$_2$/MoSe$_2$ heterobilayer without (left) and with (right) microfabrication. (d) Low-temperature PL spectrum of microfabricated WSe$_2$/MoSe$_2$ heterobilayer with various pillar sizes of 1 $\mu$m (blue), 400 nm (green), and 200 nm (red), respectively.

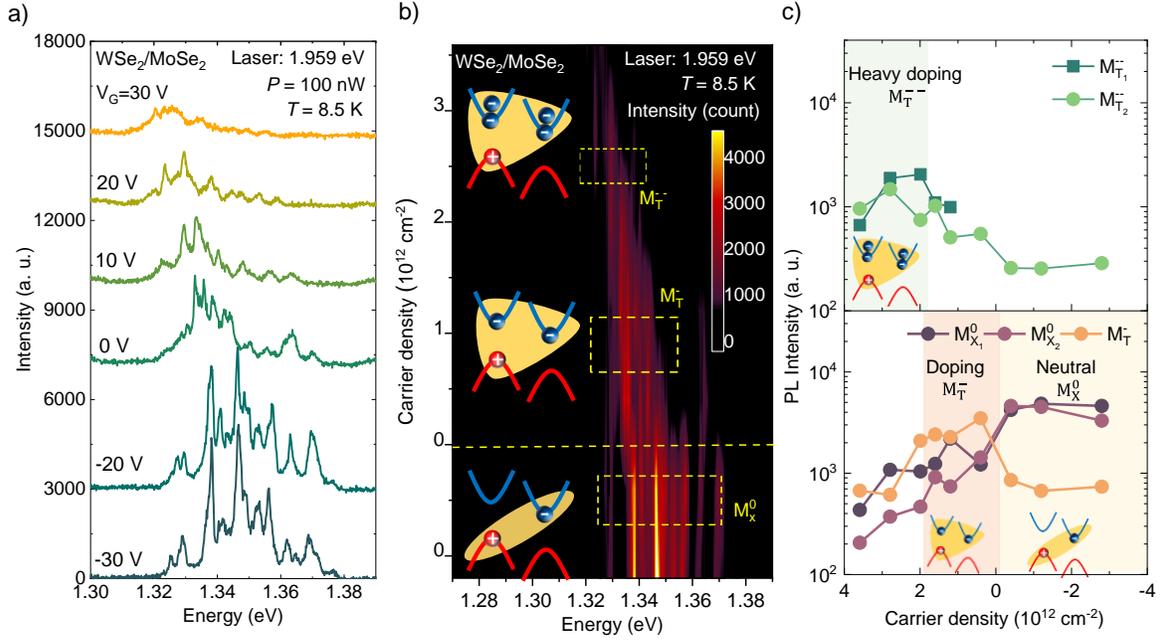

**Figure 2** **(a)** Low-temperature PL spectra of WSe$_2$/MoSe$_2$ vdW heterobilayer under excitation power condition of 100 nW and at $V_g$ ranging from −30 to 30 V. **(b)** Contour plot of PL spectra of WSe$_2$/MoSe$_2$ vdW heterobilayer as a function of doped carrier density. Schematic of electron and hole configurations. **(c)** PL intensities of traced multiple peaks monitored of neutral moiré exciton ($M_{X_1}^0$ and $M_{X_2}^0$) and charged moiré exciton or trion under slight electron-doping ($M_T^-$) and heavy electron-doping ($M_{T_1}^{--}$ and $M_{T_2}^{--}$) as a function of $V_g$.

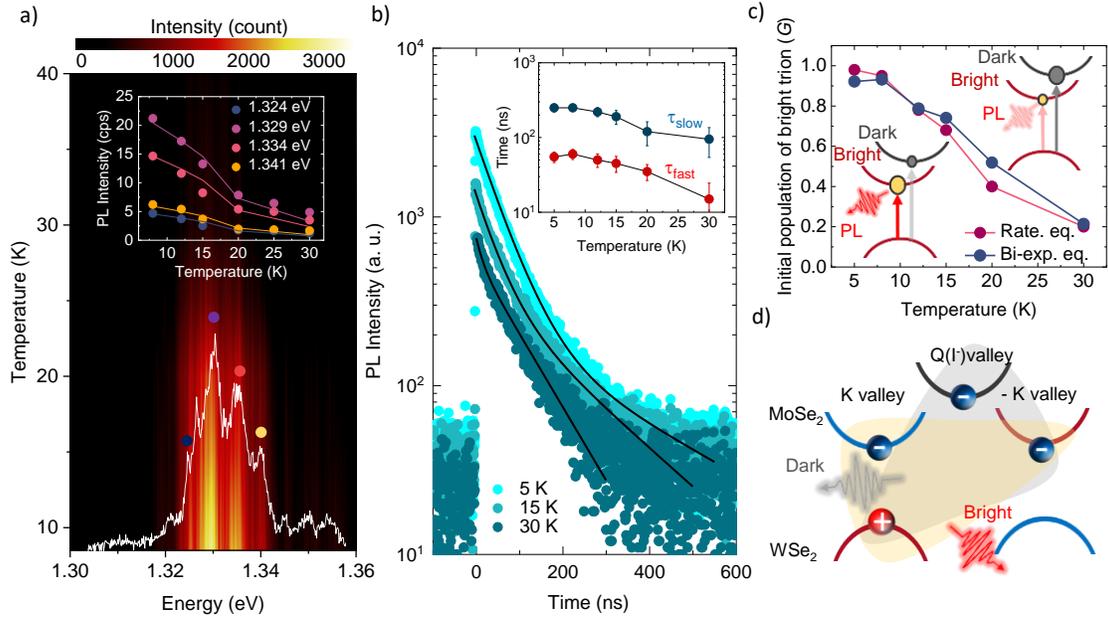

**Figure 3 (a)** Contour plot of PL spectra of WSe$_2$/MoSe$_2$ vdW heterobilayer at $V_g$ of 10 V as a function of temperature. The inset figure shows the PL intensities of traced multiple peaks monitored at 1.324, 1.329, 1.334, and 1.341 eV. **(b)** PL decay curves from 5 to 30 K under excitation power of 500 nW. The inset of the figure shows the fast ($\tau_f$, red) and slow decay components ($\tau_s$, blue) in the range of various temperatures, obtained from bi-exponential function. **(c)** Initial trion population in the bright state, $G$ from rate equation (red) and bi-exponential function (blue) as a function of temperature. **(d)** Schematic of moiré trion structures modeled with bright (K-K) and dark (Q(Γ)-K) moiré trion states.

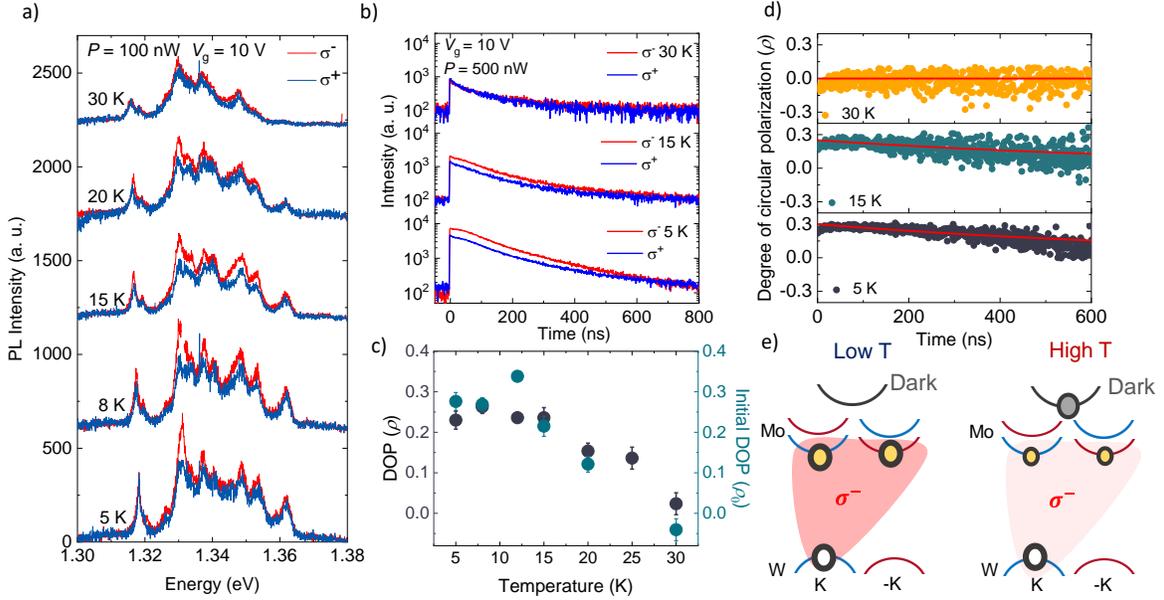

**Figure 4 (a)** Polarization-resolved PL spectra of moiré trion at various temperatures under $\sigma^+$ excitation. The excitation photon energy of 1.72 eV corresponds to the resonant excitation of exciton at the K valley in the WSe$_2$ layer of the heterobilayer. **(b)** Polarization-resolved PL decays of $\sigma^+$ and $\sigma^-$ components at 1.329 eV under $\sigma^+$ pulse excitation at various temperatures. **(c)** Degree of circular polarization $\rho$ as a function of temperature from polarization-resolved PL spectra. **(d)** Time evolution of valley polarization from the polarization-resolved PL decay curves of $I_{\sigma^+}(t)$ and $I_{\sigma^-}(t)$ at $V_g$ of 10 V from 5 to 30 K. **(e)** Schematic of bright moiré trion emission ($\sigma^-$) with different initial populations in the bright and dark states at low and high temperatures.

Supporting Information for

# Dynamics of moiré trion and its valley polarization in microfabricated WSe$_2$/MoSe$_2$ heterobilayer

*Heejun Kim[1,\*], Duanfei Dong[1,\*], Yuki Okamura[1,\*], Keisuke Shinokita[1],*

*Kenji Watanabe[2], Takashi Taniguchi[3], and Kazunari Matsuda[1]*

[1]Institute of Advanced Energy, Kyoto University, Uji, Kyoto 611-0011, Japan

[2]Research Center for Functional Materials, National Institute for Materials Science, 1-1 Namiki, Tsukuba, Ibaraki 305-0044, Japan

[3]International Center for Materials Nanoarchitectonics, National Institute for Materials Science, 1-1 Namiki, Tsukuba, Ibaraki 305-0044, Japan

**Materials and Methods**

**Methods**

**Sample preparation**

Monolayers of MoSe$_2$, WSe$_2$, a few layers of graphene (as an electrode), and *h*-BN layers were prepared by mechanical exfoliation. The WSe$_2$/MoSe$_2$ heterobilayer with top and bottom *h*-BN capping layers was fabricated using the dry-transfer technique using a polymer (polydimethylsiloxane (PDMS)) stamp on a 270-nm SiO$_2$/Si substrate. The residual PDMS was removed by immersing the sample in an acetone solution for 40 min and then thermally annealed at 150°C for 1 h in Ar/H$_2$ (97 %/3 %) to obtain better contact between the layers. Then, the metal mask was used to prepare the electrode pad, which was connected to the graphene electrode. The thicknesses of the MoSe$_2$ and WSe$_2$ monolayers were estimated by PL spectra and optical comparison.

**Microfabrication by focused Ga$^+$ ion beam technique**

The FIB technique was used to prepare the microfabricated WSe$_2$/MoSe$_2$ heterobilayer with various patterns using gallium ions (Ga$^+$). The microfabrication positions were fixed using a scanning ion microscope, and the etching process was achieved by sputtering the sample surface with Ga$^+$ ion beam irradiation. The experimental acceleration voltage is 10 kV with the electricity of 20 pA, the field of view range is approximately 2 μm, and the machining depth is set to 0.1 μm.

**Steady-state PL measurements**

Linearly polarized continuous-wave semiconductor lasers with a photon energy of 1.959 and 2.33 eV were used as excitation light sources for the PL measurement under temperature- and excitation power-dependent conditions. The laser was focused onto the sample surface using a 50× objective lens (spot size: ~1.5 μm). The emission from the $WSe_2/MoSe_2$ heterobilayer was collected by a cooled charge-coupled device through a spectrometer with a spectral resolution of 0.5 meV. The samples were cooled using a cryogen-free cryostat from 300 to 5 K. A pulsed supercontinuum light source with a photon energy of 1.72 eV (pulse duration: ~100 ps; repetition rate of 1 MHz) at all temperatures and under power-dependent conditions was used for the polarization-resolved PL measurement.

**Time-resolved PL measurements**

A pulsed supercontinuum light source with a photon energy of 1.72 eV (pulse width of approximately ~100 ps; repetition rate of 1 MHz) was used to perform the time-resolved PL measurement. A bandpass filter of $1.333 \pm 0.007$ eV was used to obtain the decay signal from the spectrum.

**Measurement of twist angle in a twisted $WSe_2/MoSe_2$ heterobilayer**

The twist angle between the monolayers of $MoSe_2$ and $WSe_2$ in the form of the heterobilayer was confirmed through optical microscope image and second-harmonic generation (SHG) measurements. From the optical microscope image, we can estimate the twist angle between two monolayers by sharp edge along the zigzag direction from the monolayer flake. In order to confirm the twist angle, we further use the linearly polarized femtosecond laser pulse (80 MHz, 900 nm, approximately 100 fs) for SHG measurement. The laser purse was focused on each WSe$_2$/MoSe$_2$ heterobilayer, i.e., the MoSe$_2$ and WSe$_2$ monolayers, and the angle dependence of each SHG signal was measured by rotating the polarization with a half-wave plate. The maximum SHG signal was obtained when the laser polarization was parallel to the zigzag directions of the monolayers and heterobilayer.

**Measurement of the charge carrier doping using capacitance model**

The carrier density was calculated using the capacitance model in the device structure, as shown in Figure S3a. The doped carrier density ($n_e$) according to the relationship between geometric capacitance and the back-gate voltage $V_g$ can be expressed as follows,[1,2]:

$$n_e = |V_g - V_0|\varepsilon\varepsilon_0/t_{SiO_2}, \tag{S1}$$

where $V_0$ denotes the back-gate voltage at the charge neutrality point in this device (−20 to −10 V). $\varepsilon$ (=3.9) denotes the dielectric constant of silicon dioxide, $\varepsilon_0$ denotes the dielectric constant of vacuum, and $t_{SiO_2}$ denotes the thickness of silicon dioxide as 270 nm. Figure 2a shows a counter map of PL spectra as a function of the doped carrier density from 0 to approximately $3 \times 10^{12}$ cm$^{-2}$, calculated from the back-gate voltage.

**Estimation of the exciton density in moiré potential by lattice constant**

The number density of the moiré potential was estimated using the lattice mismatch $\delta$ between MoSe$_2$ ($a_{MoSe_2}$= 3.29 Å) and Wse$_2$ ($a_{Wse_2}$= 3.28 Å), respectively, as follows,

$$\delta = |a_{MoSe_2} - a_{Wse_2}|/a_{MoSe_2}. \tag{S2}$$

The moiré periodicity of the moiré potential mainly determined by $\delta$ can be written as $a_M \approx a/\sqrt{\delta^2 + \theta^2}$ when lattice mismatch $\delta$ is larger than the averaged lattice constant $a$ between the two layers with a small twist angle $\theta$. The moiré periodicity shows 6–8 nm in the heterobilayer with a twist angle of 2°–3°.

**Steady-state PL intensity as a function of temperature based on rate equation analysis**

The temperature-dependent PL intensity of trapped trions within the moiré potential can be derived from the rate equation as follows:

$$\frac{dN_B}{dt} = G - N_B(\tau_{Br}^{-1} + \tau_{BD}^{-1}) + N_D(\tau_{DB}^{-1}), \tag{S3}$$

$$\frac{dN_D}{dt} = -N_D(\tau_{Dnr}^{-1} + \tau_{DB}^{-1}) + N_B(\tau_{BD}^{-1}), \tag{S4}$$

where $G$ denotes the generation rate of a trapped trion; $N_B$ and $N_D$ denote the time-dependent trion populations in the bright and dark moiré trion states, respectively; $\tau_{Br}^{-1}$ and $\tau_{Dnr}^{-1}$ denote the radiative and non-radiative decay rates and transition rates between the bright to dark ($\tau_{BD}^{-1}$) and dark to bright ($\tau_{DB}^{-1}$) states, respectively. The rate equations are solved under steady-state conditions. Then, the population of trions in the dark state can be expressed as follows.

$$N_D = \frac{N_B(T)\tau_{BD}^{-1}}{(\tau_{Dnr}^{-1} + \tau_{DB}^{-1})}. \tag{S5}$$

The temperature-dependent PL intensity $I(T)$ can be expressed under steady conditions using Eq. (S1), (S2), and (S3) as follows.

$$I(T) \propto \tau_{Br}^{-1} N_B = \tau_{Br}^{-1} \frac{G}{(\tau_{Br}^{-1} + \tau_{BD}^{-1}) - \left\{\frac{\tau_{DB}^{-1} \times \tau_{BD}^{-1}}{(\tau_{Dnr}^{-1} + \tau_{DB}^{-1})}\right\}}. \tag{S6}$$

## S1: Stacking angle of WSe$_2$/MoSe$_2$ heterobilayer using SHG measurement

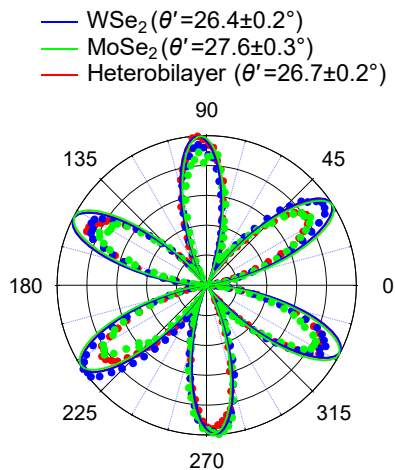

**Figure S1** Polar plot of normalized SHG intensity in 1L-WSe$_2$ (blue), 1L-MoSe$_2$ (green), and MoSe$_2$/WSe$_2$ heterobilayer (red) of sample A, respectively. The solid lines indicate the fitted results with six-fold symmetry in SHG intensity to determine the angle. The twist angle of MoSe$_2$/WSe$_2$ heterobilayer is evaluated as 1.5°, which is well consistent with the value from the optical image in Figure 1b.

## S2: Optical characteristics of microfabricated WSe$_2$/MoSe$_2$ heterobilayer

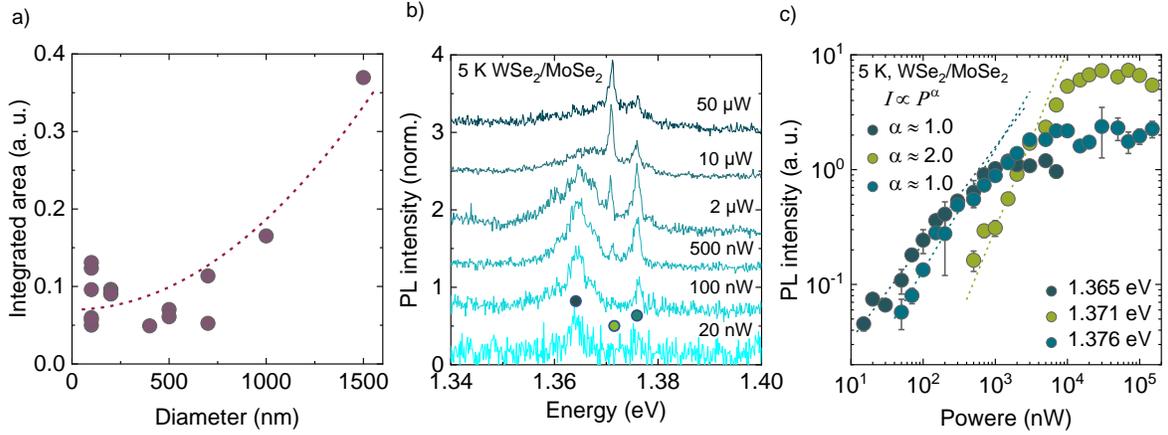

**Figure S2 (a)** Integrated PL intensity as a function of pillar diameter in microfabricated heterobilayer. The dotted line indicates the guideline for the integrated area using pillar diameter $D$, **(b)** PL spectra of WSe$_2$/MoSe$_2$ heterobilayer with $D = \sim250$ nm under various excitation power. **(c)** Integrated PL intensities of peaks at 1.365 (dark green), 1.371 (light green), and 1.376 eV (deep dark green) as a function of excitation power.

To clarify the reduced peaks from the microfabricated WSe$_2$/MoSe$_2$ heterobilayer, the integrated areas of the peaks in the heterobilayers with different pillar diameters $D$ are shown in Figure S2 (a). We observed that the integrated area reduces with a smaller pillar diameter, which implies an effective approach for reducing the number of peaks using the FIB technique. With the advantage of this approach, we were also able to explore the biexciton emission within the moiré potential from the microfabricated WSe$_2$/MoSe$_2$ heterobilayer. Figure S2(b) shows normalized PL spectra under various excitation powers, where the spectra are normalized by the excitation power. The lowest energy PL peak at 1.365 eV shows strong saturation behavior, which is also confirmed by the plot shown in Figure S2(c). The PL peak at 1.371 eV appears above the excitation power condition of 500 nW and exhibits sharp increases in the plot in Figure S2(c) with increasing excitation power, where the power law component exhibits the highly nonlinear behaviors. These are consistent with the optical characteristics of biexciton emission within the moiré potential.[3]

## S3: Optical characteristics of microfabricated WSe$_2$/MoSe$_2$ heterobilayer with graphene electrode

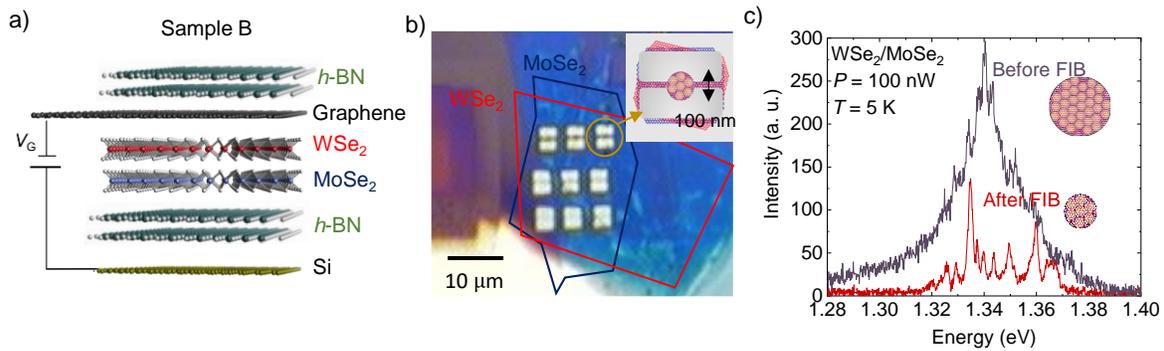

**Figure S3 (a)** Schematic of 1L- WSe$_2$/MoSe$_2$ vdW heterobilayer with top and bottom of *h*-BN and graphene electrode. **(b)** Optical image of microfabricated WSe$_2$/MoSe$_2$ heterobilayer with graphene electrode. The red and dark blue lines indicate the monolayers of WSe$_2$ and MoSe$_2$, respectively. The inset figure shows a schematic of microfabrication pattern using FIB. The thin lines of ~100-nm width indicate mechanical support and electrical line to apply the back-gate voltage in the heterobilayer. **(c)** Low-temperature PL spectrum of the heterobilayer under excitation power condition of 100 nW from different laser spot areas, before (black) and after (red) FIB process.

## S4: PL spectra of WSe$_2$/MoSe$_2$ heterobilayer under different back-gate voltage conditions

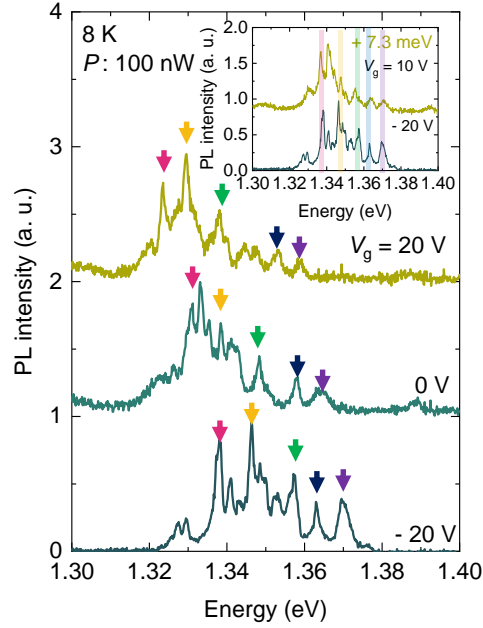

**Figure S4** PL of WSe$_2$/MoSe$_2$ heterobilayer at different back-gate voltages, where PL spectra originate from moiré exciton ($M_X^0$), and moiré trions ($M_T^-$ and $M_T^{--}$) under the neutral (deep dark green), slight electron-doping (dark green), and heavy electron-doping (light green) conditions at −20, 0, and 20 V, respectively. The inset shows the PL spectra of neutral moiré exciton ($M_X^0$) and moiré trions ($M_T^-$), where the PL spectrum of $M_T^-$ is shifted to the higher energy side of 7.3 meV.

## S5: Power-dependent PL spectra of WSe$_2$/MoSe$_2$ heterobilayer at different $V_g$

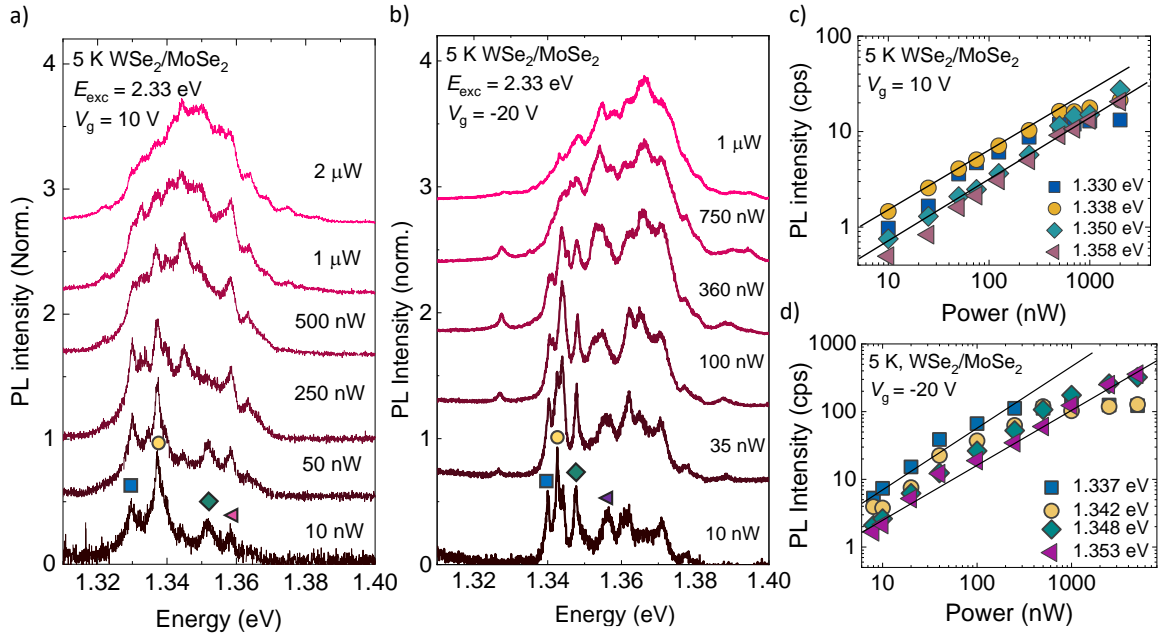

**Figure S5 (a)** and **(b)** PL spectra of WSe$_2$/MoSe$_2$ heterobilayer at various excitation power from line cut of the contour plot at $V_g$ = 10 and −20 V, respectively, where each spectrum is normalized by excitation power. (c) and (d) PL intensities monitored at various photon energies as a function of excitation power at $V_g$ = 10 and −20 V, respectively.

## S6: PL decay curves of WSe$_2$/MoSe$_2$ heterobilayer at various temperatures

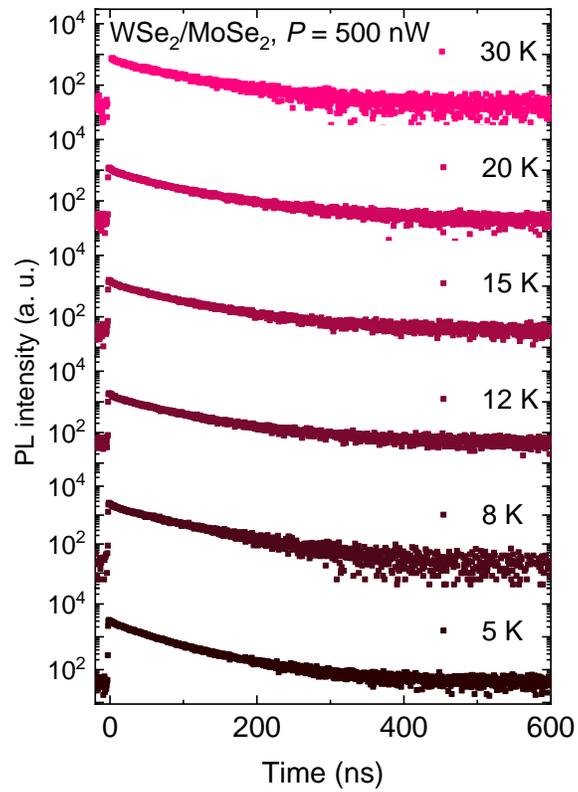

**Figure S6** PL decay curves of WSe$_2$/MoSe$_2$ heterobilayer at back-gate voltage of 10 V from 5 to 30 K. PL decay profile at 1.329 eV under excitation power of approximately 500 nW.

## S7: Analysis of temperature dependence of lifetimes and relative amplitudes of PL decay curves

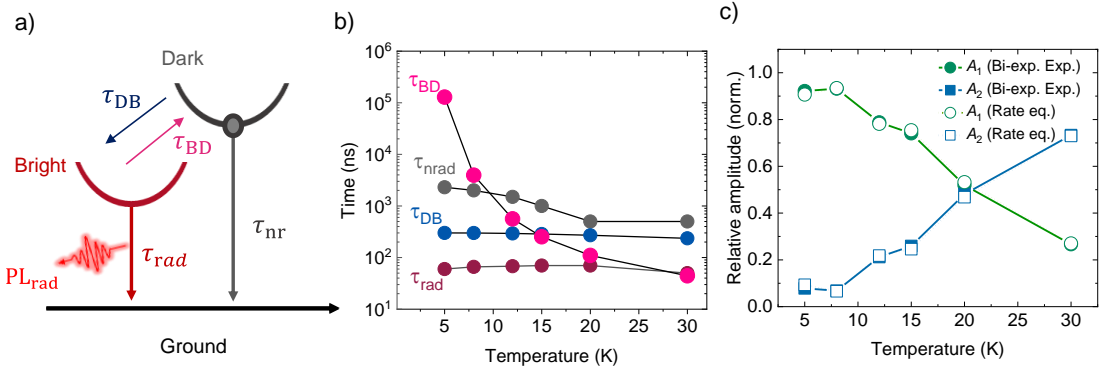

**Figure S7 (a)** Three-level energy system, including dark and bright trion and ground states. **(b)** Temperature dependence of the radiative lifetime of bright trion ($\tau_{Br}^{-1}$, red), the non-radiative lifetime of dark trion ($\tau_{Dnr}^{-1}$, gray), and transition rate of bright to dark ($\tau_{BD}^{-1}$, light red) and dark to bright ($\tau_{DB}^{-1}$, dark blue) states in the moiré potential, derived from the rate equation analysis. **(c)** Temperature-dependent relative amplitudes ($A_1$ and $A_2$) obtained from fitting procedures with bi-exponential function and calculated values using the obtained parameters from rate equation analysis.

**S8: Temperature dependence of polarization-resolved PL decays and transient degree of valley polarization**

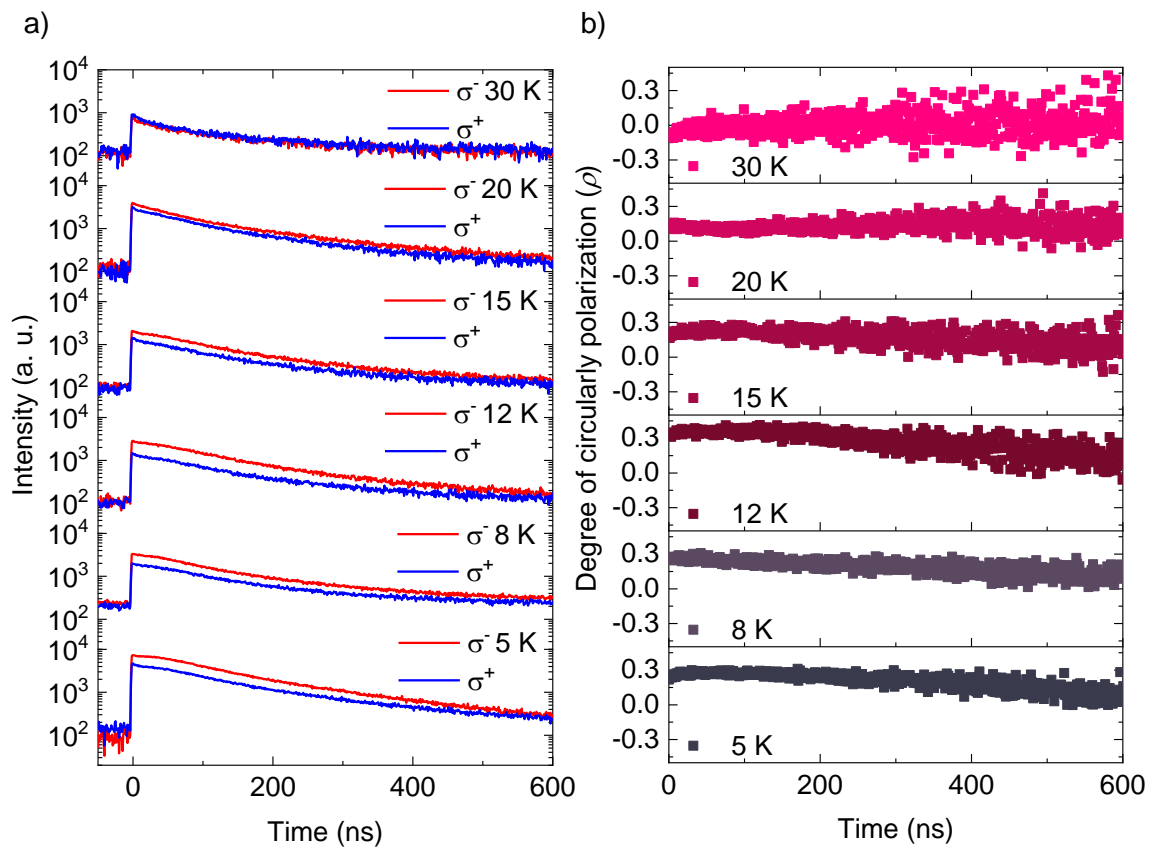

**Figure S8 (a)** Temperature dependence of polarization-resolved PL decay with $\sigma^+$ (blue) and $\sigma^-$ (red) components at 1.329 eV under $\sigma^+$ excitation, **(b)** Time evolution of the degree of circular polarization as a function of temperature.

**S9: Temperature dependence of valley polarization from polarization-resolved PL decay curves**

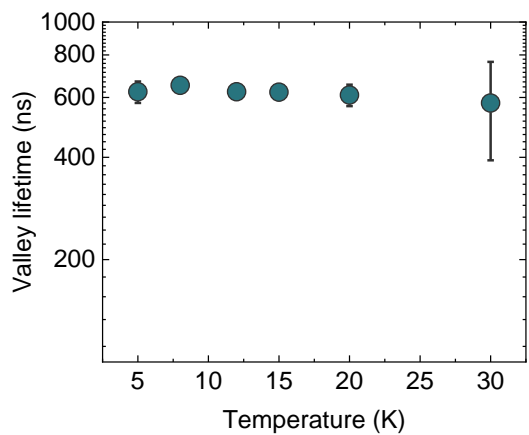

**Figure S8** Temperature dependence of valley polarization from the polarization-resolved PL decay curves, fitted by a single exponential function.

**Reference**

1. Tang, Y. *et al.* Tuning layer-hybridized moiré excitons by the quantum-confined Stark effect. *Nat*. *Nanotechnol* **16**, 52–57 (2021).

2. Sung, J. *et al.* Broken mirror symmetry in excitonic response of reconstructed domains in twisted MoSe$_2$/MoSe$_2$ bilayers. *Nat*. *Nanotechnol* **15**, 750–754 (2020).

3. Li, W. *et al*. Dipolar interactions between localized interlayer excitons in van der Waals heterostructures. *Nat*. *Mater* **19**, 624–629 (2020).